\documentclass[pre,aps,amsmath,amsfonts,amssymb,twocolumn]{revtex4-1}
\usepackage{graphicx,bm}
\begin{document} 
\title{Critical dynamics of nonconserved $N$-vector model with anisotropic nonequilibrium perturbations}
\author{Sreedhar B. Dutta}
\affiliation{School of Physics, Indian Institute of Science Education and Research, Thiruvananthapuram, India}
\author{Su-Chan Park}
\email{spark0@catholic.ac.kr}
\affiliation{Department of Physics, The Catholic Univeristy of Korea, Bucheon 420-743, Korea}
\date{\today}
\begin{abstract} 
We study dynamic field theories for nonconserving $N$-vector models that are 
subject to spatial-anisotropic bias perturbations. We first investigate the 
conditions under which these field theories can have a single length scale.
When $N=2$ or $N \ge 4$, it turns out that there are no such field theories, and, 
hence, the corresponding models  are pushed by
the bias into the Ising class.
We further construct nontrivial field theories for $N=3$ case with certain bias perturbations 
and analyze the renormalization-group flow equations. We find that the three-component systems can exhibit rich critical behavior 
belonging to two different universality classes.
\end{abstract}
\maketitle
\section{Introduction}
Classification of the universality exhibited by systems with macroscopic 
degrees of freedom, both at and away from equilibrium, is one of the main 
objectives that has been pursued in statistical physics ever since the advent 
of scaling theory and renormalization-group (RG) framework.  The universality 
classes of nonequilibrium systems are far less understood, unlike those at 
equilibrium, in spite of having identified many nonequilibrium 
classes such as the absorbing phase transitions~\cite{MD1999}, growing 
surfaces~\cite{BS1995}, self-organized criticality~\cite{B1996}, 
driven diffusive systems~\cite{SZ1995}, and so on. 

Constructing classes of infrared-stable field theories by taking a scaling 
limit of microscopic models is a formidable task, even at equilibrium. Hence, 
probing known field theories by various perturbations and following the 
induced instabilities, if any, is an alternative that can provide invaluable 
insights towards any  classification.

Near-equilibrium critical dynamics is extensively studied and effectively 
captured by time-dependent Landau-Ginzburg (LG) models as categorized by 
Hohenberg and Halperin~\cite{hohenberg}. Recent studies have explored the 
effects of nonequilibrium perturbations  on various dynamic universality 
classes~\cite{grinstein, bassler, tauber, akkineni, tauber-2002, jayajit}.  
They not only include perturbations about the LG energy functionals but also 
{\it genuine} nonequilibrium perturbations about the critical dynamics. 
The detailed-balance violating perturbations turn out to be relevant in the
conserved systems~\cite{SZ1995,tauber,BR1995}. On the other hand,
it is well established that the kinetic Ising systems of Model-A
class (in Hohenberg-Halperin classification) are stable against local dynamic
perturbations, even if they violate detailed-balance condition, provided the
symmetries are preserved~\cite{grinstein, haake-1984}. Bassler and
Schmittmann (BS) further found that the spatially anisotropic perturbations, in
spite of not respecting the $Z_2$ symmetry, cannot destabilize the dynamic
class of nonconserved kinetic Ising models, which are described by a single
scalar order-parameter field~\cite{bassler}.
This naturally brings forth the issue whether the irrelevance of such 
spatially anisotropic
perturbations pervades throughout Model-A systems or is only restricted to 
its subset, like those describable by a scalar order parameter. 
It was presumed that the $N$-component systems, such as Kinetic Ising models, might also be robust to such perturbations~\cite{odor,tauber}. 
We find that, upon investigating the role of spatially anistropic  
perturbations on $N$-component Model-A systems, that this is not the case.

The structure of this paper is as follows: In Sec.~\ref{Sec:RG},
we construct the $N$-component Model-A system with anisotropic non-equilibrium 
perturbations, and then address the possibility of constructing a field theory with a single characteristic length scale. We show that, unless $N=3$, 
the system should belong to the Ising class, which is confirmed
numerically for the case of $N=2$. 
In Sec.~\ref{Sec:N=3},  we analyze $N=3$ systems using the renormalization-group techniques. 
In Sec.~\ref{Sec:sum}, we summarize the results.

\section{\label{Sec:RG} Permutation-symmetric $N$-vector dynamic critical field theories}

In this section, we construct nonconserving $N$-vector models subject
to spatial-anisotropic perturbations and find the interactions consistent 
with a single length scale. 

We consider the following class of $N$-vector models driven by a nonconserved Langevin dynamics:
\begin{equation}
\label{laneqn}
{\partial}_t\phi_a(\bm{x},t) =  {\cal F}_a(\phi(\bm{x},t)) + \eta_a(\bm{x},t), 
\end{equation}
with
\begin{eqnarray}
{\cal F}_a(\phi) =  (\nabla^2 - r)\phi_a 
+\frac{{\cal E}_{abc}}{2}  \phi_b \partial_\| \phi_c
- \frac{G_{abcd}}{3!} \phi_b\phi_c\phi_d,
\end{eqnarray}
where the indices $a,b,c$ and $d$ run from $1$ to $N$, the summation over repeated
indices is assumed,
and $\eta_a(\bm{x},t)$ denotes the Gaussian noise with zero mean and variance 
$\langle \eta_a(\bm{x},t)\eta_b(\bm{x}',t')\rangle= 
2 T \delta_{ab}\delta(\bm{x}-\bm{x}')\delta(t-t')$.  
Since $\phi_b \phi_c \phi_d = \phi_d \phi_b \phi_c$ and so on, we assume that,
without any loss of generality,  $G_{abcd}$ is invariant under all permutations of $\{b,c,d\}$ ($G_{abcd} = G_{adbc}$, for example).  The couplings ${\cal E}_{abc}$ introduce spatial anisotropy in the $x_{\parallel}$ direction. The 
spatial-anisotropic perturbations, often referred to as the bias, are straightforward generalizations of the bias perturbation in the BS model~\cite{bassler}. Note that the above $G$ and ${\cal E}$ interaction terms are the most general marginal perturbations at d = 4.

It should be remarked that, if ${\cal F}_a(\phi)$ is derivable from a functional $S[\phi]$, namely,  
${\cal F}_a(\phi)(x)= - \delta S[\phi] / \delta \phi_a(x)$,  then (under certain conditions) the system exhibits equilibrium behaviour at large times. 
Any term that is not derivable from a functional when included will not allow the system to equilibrate; hence, it shall be referred to as {\it genuine} non-equilibrium perturbation.  Unlike most of the $G$ terms, the ${\cal E}$ terms are genuine nonequilibrium perturbations and can lead the system to a variety of nonequilibrium states.

We  now investigate which of the interactions are consistent with a field theory with 
a single characteristic length scale in the long-time limit. 
We shall find such interactions by first demanding  that the set of equations~\eqref{laneqn} are  
invariant under any permutation of the field components, and then demanding 
the existence of a single length-scale. 

\subsection{Permutation-symmetric  interactions}
Let $\hat P$ be an operator transforming
Langevin equations such a way that $\hat P \partial_t \phi_a \equiv \partial_t
\phi_{{\cal P}a}$ and
\begin{eqnarray}
\hat P {\cal F}_a(\phi) &\equiv&  (\nabla^2 - r)\phi_{{\cal P}a} 
+ \frac{1}{2}\sum_{bc}{\cal E}_{abc} \phi_{{\cal P}b} \partial_\| \phi_{{\cal P}c} \nonumber\\
\nonumber\\
&-& \frac{1}{3!}\sum_{bcd} G_{abcd} \phi_{{\cal P}b}\phi_{{\cal P}c}\phi_{{\cal P}d} \nonumber\\
 &=&  (\nabla^2 - r)\phi_{{\cal P}a} 
+ \frac{1}{2}\sum_{bc}{\cal E}_{a{\cal P}^{-1}b{\cal P}^{-1}c} \phi_{b} \partial_\| \phi_{c}
\nonumber\\
&-& \frac{1}{3!}\sum_{bcd} G_{a{\cal P}^{-1}b{\cal P}^{-1}c{\cal P}^{-1}d} \phi_{b}\phi_{c}\phi_{d},
\end{eqnarray}
where ${\cal P}$ is a permutation of field components $\{1,\ldots,N\}
\mapsto \{ {\cal P}1, \ldots, {\cal P}N\}$ with ${\cal P}^{-1}$ to be its inverse.
Since a permutation-symmetric theory demands that Eq.~\eqref{laneqn} should be
invariant under $\hat P$, that is, $\hat P {\cal F}_a = 
{\cal F}_{{\cal P}a}$, the coupling constants should satisfy
${\cal E}_{{\cal P}a bc} = {\cal E}_{a {\cal P}^{-1} b {\cal P}^{-1} c}$
and $G_{{\cal P} a bcd} = G_{a{\cal P}^{-1} b {\cal P}^{-1} c {\cal P}^{-1} d}
$ or, equivalently,
\begin{equation}
{\cal E}_{abc} = {\cal E}_{{\cal P}a {\cal P}b{\cal P}c},\quad
G_{abcd} = G_{{\cal P} a {\cal P}b{\cal P}c{\cal P}d},
\end{equation}
for all ${\cal P}$'s and $a,b,c$ and $d$.

The permutation symmetry in the dynamics will restrict the number of 
independent $G$ couplings to seven, which are denoted as 
\begin{equation}
G_{1111}, G_{1112}, G_{1122}, G_{1123}, G_{1222}, G_{1223}, G_{1234} . 
\label{Eq:G_inde}
\end{equation}
The notation $G_{1111}$ refers to those couplings 
$G_{abcd}$, where all the indices $b,c$ and $d$ are same 
as $a$, and  $G_{1112}$ is used when one of the indices $b,c$ and $d$ is 
different from $a$, and so on. 
Recall that, by construction, 
$G_{abcd}$ is assumed to be invariant under all permutations in
$\{b,c,d\}$.
If any of the indices of a coupling constant is greater than $N$, then that 
coupling constant is understood to be zero. 
Likewise, there are five allowed bias couplings:
\begin{equation}
\label{E-coup}
 {\cal E}_{111}, {\cal E}_{112}, {\cal E}_{121}, {\cal E}_{122}, {\cal E}_{123}.
\end{equation}
Although the permutation symmetry does not require ${\cal E}_{112} = {\cal E}_{121}$, it does demand ${\cal E}_{123}= {\cal E}_{132}$.

Note that if we soften the permutation symmetry to cyclic-permutation symmetry, then there are more number of allowed coupling constants. We shall later consider dynamic models with only the cyclic-permutation symmetry.

\subsection{\label{Sec:single}Interactions consistent with a single length scale}

In order to identify the couplings that are consistent with a single length scale (or mass scale), it is convenient to analyze Eq.~\eqref{laneqn} in Martin-Siggia-Rose (MSR) formalism~\cite{martin}. 
The MSR action for Eq.~\eqref{laneqn} is given by 
\begin{eqnarray}
&&{\cal S}(\widetilde{\phi}, \phi) =
\int_{x} 
\left[ \widetilde{\phi}_a \Big( \partial_t \phi_a - {\cal F}_a(\phi) \Big) 
- T \widetilde{\phi}_a \widetilde{\phi}_a 
\right]\nonumber\\
&&= \int_{x}\Bigl [
\widetilde \phi_a (\partial_t - \nabla^2 + r ) \phi_a 
-\frac{1}{2} {\cal E}_{abc} \widetilde \phi_a \phi_b \partial_\|\phi_c 
+ \nonumber\\
&&+ \frac{1}{3!} G_{abcd} \widetilde \phi_a \phi_b \phi_c\phi_d 
- T \widetilde \phi_a\widetilde \phi_a \Bigr ],
\label{msr}
\end{eqnarray}
where $\int_x \equiv \int dt d^d \bm{x}$,  
$\widetilde{\phi}_a$ refers to the auxiliary (response) field, the 
conventions $\phi_a = \phi_a(\bm{x},t)$ and $\widetilde{\phi}_a=
\widetilde{\phi}_a(\bm{x},t)$ are used, and the summation over repeated 
indices is assumed.

The permutation symmetry in the above-constructed MSR action~\eqref{msr}  with seven $G$-couplings and five ${\cal E}$-couplings
is only a necessary condition for single length scale (or mass scale). However, it is not sufficient since  
there are other relevant terms allowed by the symmetry that may get generated during renormalization, such as
$\widetilde{\phi}_a \widetilde{\phi}_b$,  $\widetilde{\phi}_a {\phi_b}$, and
$\widetilde{\phi}_a \partial_{\parallel}^2 {\phi_b}$, where $a\!\neq \! b$.
In particular, it is the off-diagonal mass term $\sum_{a\neq b} 
\widetilde{\phi}_a M_{ab} {\phi_b}$, if generated, that will introduce an 
extra length scale. In fact, the permutation symmetry will imply that all the diagonal elements are equal and, similarly,
all the off-diagonal elements are equal. This mass matrix will have two eigenvalues, one of which is $N\!-\!1$ degenerate. Therefore, the
presence of off-diagonal mass-terms in an $N$-vector model indicates  a crossover of the critical behavior to either that of a scalar model or to that of a $N\!-\!1$-vector model (which itself may not have a single length scale).

Now the question boils down to which form of the interactions will avoid the generation of the
off-diagonal mass  during renormalization. Before we present more general symmetry arguments for identifying those interactions, we shall specify the conditions that are imposed by perturbative corrections to second order.

At  one loop,  the ${\cal E}$ couplings may generate off-diagonal kinetic terms $ \widetilde \phi_a \partial^2_{\parallel}\phi_b$, and the $G$ couplings
may generate off-diagonal mass term proportional to $\sum_{c} G_{abcc} \widetilde \phi_a \phi_b $.  The off-diagonal mass terms are absent only if the coupling constants satisfy the  {\it trace condition}~\cite{brezin}: $\sum_{c} G_{abcc}=0$ for $a \ne b$ and $N \ge 2$, which, when expressed explicitly, is
\begin{equation}
G_{1112} + G_{1222} + (N-2)G_{1223} =0.
\end{equation}
Provided the ${\cal E}$ couplings have generated nonzero off-diagonal kinetic term at one loop, then the two-loop corrections to the off-diagonal mass are  proportional to $\sum_{c,d} G_{abcd} \widetilde \phi_a \phi_b $. Hence, for the absence of off-diagonal mass terms, the coupling constants need to satisfy a further trace condition,
 \begin{equation}
2G_{1122} + 2(N-2) (G_{1123} + G_{1223} )+ (N-2)(N-3)G_{1234} =0.
\end{equation}

Finding further constraints from higher-order correction is rather cumbersome. Instead, we invoke
symmetry arguments to find the coupling constants that are consistent with a single length scale. 
To this end, we define certain parity symmetries and then explain how these symmetries can distinguish the presence or absence of off-diagonal mass. To any finite order, the effective action will contain terms of the form ($\tilde n_a, n_a \ge 0$)
\begin{equation}
\prod_{a=1}^N \left ( \widetilde \phi_a \right )^{\tilde n_a} 
\left ( \phi_a \right )^{n_a},
\end{equation} 
suppressing the possible derivatives.  
If $(n_a +\tilde n_a) - (n_b + \tilde n_b)$ is even
for any pair of $a,b$, we will define this term as parity symmetric.
If a term is parity symmetric and, further, $n_1+\tilde n_1$ is even (odd),
this term is said to be even(odd) parity symmetric.
Note that diagonal mass terms are even parity symmetric and off-diagonal
mass terms are not parity symmetric unless $N=2$, in which case they are odd parity symmetric.
Essentially, the diagonal mass terms have different symmetry from the off-diagonal terms. 
In the case of $N\!>\!2$, if 
the (bare) action contains interaction terms that are not parity symmetric, then the off-diagonal mass terms should emerge during renormalization;
in the case $N\!=\!2$, the odd-parity-symmetric interactions will also generate off-diagonal mass terms during renormalization.

It is easy to check that for arbitrary $N$  the terms associated with  $G_{1111}$ and $G_{1122}$ 
are even parity symmetric, while those combined with $G_{1123}$ and $G_{1223}$ 
are not parity symmetric. The couplings $G_{1112}$ and $G_{1222}$ generate terms that are
not parity symmetric for $N>2$, while they generate odd parity symmetric terms including off-diagonal mass for $N=2$. Hence, the presence
of any of the four couplings $G_{1123}$, $G_{1223}$, $G_{1112}$,  and $G_{1222}$ will generate off-diagonal mass terms and these terms should be dropped 
in order to construct a field theory with single length scale.
The coupling $G_{1234}$  is odd parity symmetric for $N=4$, while it will generate terms that are not parity symmetric for $N>4$.  Hence, $G_{1234}$ introduces off-diagonal mass for any $N>4$, but not for $N=4$. 
We shall not pursue further the $N=4$ case, since it is not relevant for the effects of spatial anisotropy,

Similarly, in multicomponent models with bias, the couplings ${\cal E}_{111}$,  ${\cal E}_{112}$, ${\cal E}_{121}$, and ${\cal E}_{122}$
are not parity symmetric.
The coupling constant ${\cal E}_{123}$  is not parity symmetric for $N>3$, but
it becomes odd parity symmetric for $N=3$. Since the off-diagonal mass terms
are not parity symmetric for $N=3$, the symmetry embedded in ${\cal E}_{123}$ for $N=3$ does not allow for the generation of off-diagonal mass during renormalization.
Hence, ${\cal E}_{123}$ is the only coupling constant which
does not generate off-diagonal mass, and that too only when $N=3$. 
We summarize these results in Table~\ref{tab1}.

\begin{table}
\caption[tab1]{\label{tab1}Parity of terms associated with each coupling constant. N-PS refers to not parity symmetric;
E-PS and O-PS refer to even parity symmetric and odd parity symmetric, respectively. 
}
\begin{ruledtabular}
\begin{tabular}{cccc} 
$N$ &N-PS  & E-PS &  O-PS \\ \hline 
Any $N$ & $G_{1123}, G_{1223}$ & $G_{1111}, G_{1122}$&  \\ 
 $N = 2$&  &  & $G_{1112}, G_{1222}$\\ 
 $N > 2$&  $G_{1112}, G_{1222}$& & \\ 
 $N = 4$&  &  & $G_{1234}$\\ 
 $N > 4$&  $G_{1234}$& & \\ 
 $N > 1$&  ${\cal E}_{111}, {\cal E}_{112}, , {\cal E}_{121},{\cal E}_{122}$& & \\ 
 $N = 3$& & & ${\cal E}_{123}$ \\ 
 $N > 3$&  ${\cal E}_{123}$& & 
  \end{tabular}
\end{ruledtabular}
\end{table}
\begin{table}[b]
\caption[tab 2]{\label{tab2}Allowed multicomponent permutation-symmetric $N$-vector field theories with a single length scale.
}
\begin{ruledtabular}
\begin{tabular}{cc} 
Components & Allowed couplings \\ \hline 
$N = 3$& $G_{1111}, G_{1122}, {\cal E}_{123}$\\ 
 $N = 4$& $G_{1111}, G_{1122}, G_{1234}$\\
 $N=2$ or $N > 4$&  $G_{1111}, G_{1122}$
\end{tabular}
\end{ruledtabular}
\end{table}

Notice that the off-diagonal mass terms and the off-diagonal kinetic terms have the same parity-symmetry.  Therefore the coupling constants which do not generate off-diagonal mass will also not generate off-diagonal kinetic terms, and hence
the second `trace condition' is not applicable. As expected all the field theories with single length-scale satisfy the first `trace condition'.

To summarize, as shown in Table~\ref{tab2},  we find that for $N=2$ or $N \ge 4$ the only $N$-vector field theories with a single length-scale  are those which do not have any coupling constants other than $G_{1111}$ and $G_{1122}$.
In these cases, the bias perturbations will eventually make the system crossover to the single-scalar field theory with bias that is studied
in Ref.~\cite{bassler}.
In case of $N=4$, the possible single length-scale theories do not have any coupling constants other than $G_{1111}$, $G_{1122}$, and  $G_{1234}$. Only in the case of $N=3$,  it is possible to have a single-length scale model subjected to bias, where the allowed coupling constants are  $G_{1111}, G_{1122}$, and ${\cal E}_{123}$.

\subsection{Numerical study for $N=2$ with bias}

In this section, we numerically confirm that an $N=2$ model with bias crosses over to the Ising class.

Consider an $O(2)$-symmetric model on a two-dimensional lattice described by the Hamiltonian 
\begin{equation}
{\cal H} = - \sum_{\langle n,m \rangle} \vec \phi_n \cdot \vec \phi_m 
+ \sum_n \left ( \frac{r + 2 d}{2} 
\phi_n^2 + \frac{u}{4}  \left ( \phi_n^2\right )^2
\right ),
\end{equation}
where $n$ is the site index, $\langle n,m \rangle$ denotes sum over all nearest neighbor pairs, $\vec \phi_n = ( \phi_{n,1}, \phi_{n,2})$ is a real 
two-component vector field, and $\phi_n^2 := \vec \phi_n \cdot \vec \phi_n $. 
The dynamics of the field $\phi_{n,a}$ in the presence of a bias is governed by the following Langevin equation:
\begin{equation}
\frac{\partial}{\partial t} \phi_{n,a}
= -\frac{\partial {\cal H}}{\partial \phi_{n,a}}+
E \partial_\| 
\left ( \phi_{n,a}^2 \right ) + \eta_{n,a}(t),
\label{Eq:simulation}
\end{equation}
where $\eta_{n,a}$ is the white noise with correlation
$\langle \eta_{n,a}(t) \eta_{n',a'}(t')\rangle =  \delta_{nn'} \delta_{aa'}\delta(t-t')$, and
$\partial_\| \left (\phi_{n,a}^2 \right ) := \phi_{n+1,a}^2 - \phi_{n-1,a}^2$, where
$n+1$ and $n-1$ refer to the two nearest neighbors of $n$ along a specified direction. 
In the absence of the bias $E=0$, the steady state of Eq.~\eqref{Eq:simulation} is described by the  partition function 
\begin{equation}
Z = \int_{-\infty}^\infty \left ( \prod_n \prod_{a=1}^2 d \phi_{n,a} \right ) e^{- 2 {\cal H}}.
\end{equation}

We have taken the two-dimensional square lattice to be of size $L\times L$ with periodic boundary conditions. 
The values of $u$ and the bias $E$ are set to unity, i.e., $u=E=1$.  
Equation~(\ref{Eq:simulation}) is then integrated numerically by employing  the Euler method with 
$\Delta t = 0.0025$. The initial condition is taken to be $\phi_{n,a} = \delta_{a,1}$ for all realizations. 
The system sizes of $L=2^6$, $2^7$, and $2^8$ are considered, and the equilibration time is set to 20~000. After equilbration we measured the magnetization
$\vec {\cal M} = \sum_n (\phi_{n,1}, \phi_{n,2})/L^2$ as well as $M_2 := | \vec{\cal M}  |^2$ and $M_4 := M_2^2$ at every 
five unit times, namely after every 2000 iterations with the above-mentioned $\Delta t $, and then
obtained the averages for all these quantities.
The critical point $r_c$ is located using the Binder cumulant
\begin{equation}
U_L = 1 - \frac{ \langle M_4 \rangle}{3 \langle M_2 \rangle^2}.
\end{equation}
The critical exponents  $\beta$ and  $\nu$ are found from finite size scaling by taking the scaling form for $\sqrt{\langle M_2 \rangle}$ to be
\begin{equation}
\sqrt{ \langle M_2 \rangle} = L^{-\beta/\nu} f(( r_c-r) L^{1/\nu}).
\label{Eq:fss}
\end{equation}
The asymptotic behavior of the universal scaling function $f$ is given by
\begin{equation}
f(y) \rightarrow \left \{ 
\begin{matrix} y^\beta & \text{ as $y\rightarrow \infty$,}\\
(-y)^{\beta - \nu}&\text{ as $y \rightarrow -\infty$.}
\end{matrix}
\right .
\label{Eq:asym}
\end{equation}

\begin{figure}[t]
\includegraphics[width=0.45\textwidth]{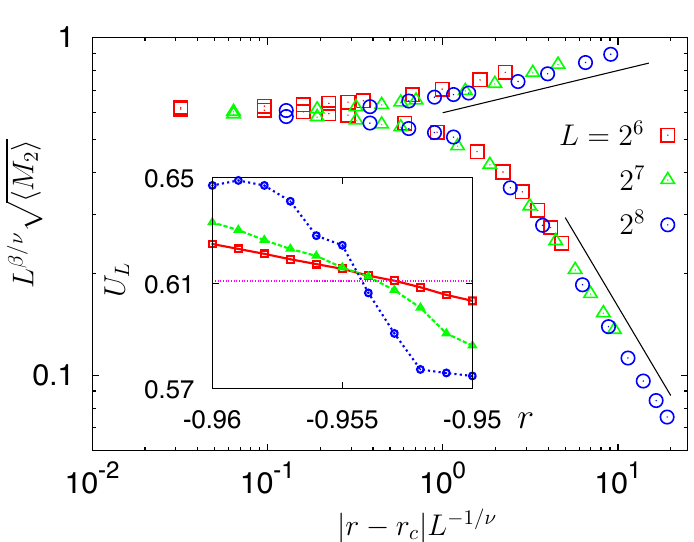}
\caption{\label{Fig:fss} (color online) Finite-size scaling collapse using 
data for $L=2^6$ (square), $2^7$ (triangle), and $2^8$ (circle). The upper (lower) 
straight line whose slope is $\frac{1}{8}$ ( $-\frac{7}{8}$) indicates the expected 
asymptotic behavior of the scaling function up to a multiplication factor. Inset: Binder
cumulants as a function of $r$ for different system sizes as in the main figure. For comparison,
the critical Binder cumulant for the Ising class is drawn as a straight line.}
\end{figure}
Numerical results are shown in Fig.~\ref{Fig:fss}. The data collapse with the asymptotic behavior
Eq.~\eqref{Eq:asym} is in good agreement with the critical exponents of two dimensional Ising model
$\beta = \frac{1}{8}$ and $\nu = 1$~\cite{Plischke}.
The critical point is located at $r_c = -0.9545 \pm 0.0010$ as shown in the inset of Fig.~\ref{Fig:fss}. The value of the critical cumulant is also consistent with that of the Ising model on a square lattice 
($\simeq 0.6107$)~\cite{KB1993}.
Thus, the model with dynamics Eq.~\eqref{Eq:simulation} clearly shows the order-disorder phase transition and exhibits critical behavior unlike its equilibrium counterpart, which can not undergo such a transition in two dimensions~\cite{mermin}.

\section{\label{Sec:N=3}Renormalization-group analysis of $N=3$ dynamic field theories with cyclic-permutation symmetry}

In Sec.~\ref{Sec:single}, we looked for permutation-symmetric $N$-vector field theories with a single length scale.
Relaxing the symmetry to {\it cyclic-permutation} symmetry can lead us to a larger set of such field theories.
In this section, we shall explore the renormalization-group fixed points of this larger set of dynamic field theories in the case $N\!=\!3$.

By cyclic-permutation symmetry for $N=3$, we mean the invariance of the MSR action
under the transformation ${\cal CP}:  1\rightarrow 2 \rightarrow 3 \rightarrow 1$.
Note that ${\cal CP}$ symmetry distinguishes $G_{1122}$ from $G_{1133}$, and furthermore allows to include the term 
$\left ( \phi_{a+2}\partial_\| \phi_{a+1} - \phi_{a+1}\partial_\| \phi_{a+2} \right )$  in ${\cal F}_a(\phi)$.
Hence, the MSR action for $N=3$ dynamic theory with ${\cal CP} $ symmetry can be written as
\begin{eqnarray}
S &=& \sum_{a=1}^3 \int_{x} \biggl [ 
\widetilde \phi_a (\partial_t - D(\nabla_\perp^2 +
\rho \partial_\|^2 - r) ) \phi_a - T \widetilde \phi_a^2+
 \nonumber\\
 &+&\sum_{i=0}^2 \frac{u_i}{3!}(3-2\delta_{i0})  \widetilde \phi_a \phi_a 
\phi_{a+i}^2
+ e_p  \phi_{a+1}\phi_{a+2}\partial_\| \widetilde \phi_a  +
\nonumber\\
&+& e_m \widetilde \phi_a   \left ( \phi_{a+2}\partial_\| \phi_{a+1} - 
\phi_{a+1}\partial_\| \phi_{a+2} \right )
\biggr ].
\label{msr3}
\end{eqnarray}
Here $D$ and $\rho$ are introduced, anticipating that these coupling constants flow separately under the RG.
The field indices take modulo-3 integer values, and hence, $\phi_4$ and $\phi_5$  mean 
$\phi_1$ and $\phi_2$, respectively. 
For notational simplicity, we relabel the couplings as  
$u_0 = G_{1111}$, $u_1 = G_{1122}$, $u_2 = G_{1133}$, and $e_p = {\cal E}_{123}$.

If we choose $u_1=u_2$ and $e_m=0$, then the action has full permutation symmetry, as 
discussed in the previous section; for the choice
$u_1=u_2= u_0/3$ and $e_m=e_p=0$, it has  $O(3)$ symmetry.
A special case, with the choice $u_1 = u_2$ and $e_p=0$, was studied in Ref.~\cite{jayajit}.

\begin{figure}[t]
\includegraphics[width=0.45\textwidth]{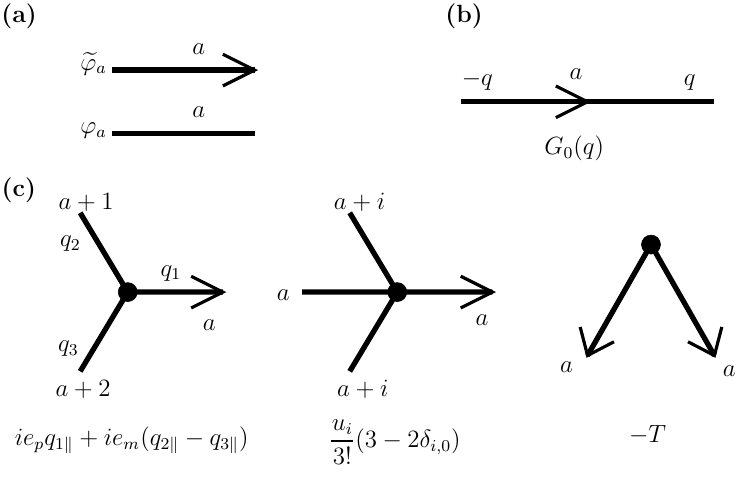}
\caption{\label{Fig:bb} Building blocks of the diagrammatic perturbations.
(a) The field with (without) a tilde in the frequency-momentum domain
is represented by a line segment
with (without) an arrow head. (b) The propagator $G_0(q)$ is
drawn using an arrow head in the middle. The four momentum  of the field
$\widetilde \varphi_a$ ($\varphi_a$) is $-q$ ($q$).
(c) Three-, four-, two-legs vertices are depicted with their interaction
strength. $a$ can be any of $\{1,2,3\}$ and $i \in \{0,1,2\}$. 
}
\end{figure}

The free theory action is given by
\begin{equation}
S_0 = \sum_{a=1}^3 \int_{q} 
\widetilde \varphi_a(-q) \left (-i \omega t + M(\bm{q}) \right ) \varphi_a(q),
\label{Eq:noninter}
\end{equation}
where the $\varphi$ are the Fourier-transformed fields 
\begin{eqnarray}
\widetilde \phi_a(\bm{x},t) = \int_q \exp\left (-i \omega t + i \bm{q} \cdot 
\bm{x} \right ) \widetilde \varphi_a(q),\\
\phi_a(\bm{x},t) = \int_q \exp\left (-i \omega t + i \bm{q} \cdot 
\bm{x} \right )  \varphi_a(q),
\end{eqnarray}
and $q$ stands for the four-momentum $(\bm{q},\omega)$;  the integral $\int_q := (2\pi)^{-(d+1)}\int dw d^d\bm{q}$; 
and 
\begin{equation}
M(\bm{q}) = D \left ( \bm{q}_\perp^2 + \rho q_\|^2 + r \right),
\end{equation}
where $q_\|$ ($\bm{q}_\perp$) denotes the component 
of $\bm{q}$  parallel (perpendicular) to the bias direction.
The  free propagator is calculated as
\begin{equation}
\langle \tilde \varphi_a(q')
\varphi_b(q) \rangle_0 
= \frac{ \delta_{ab}
\bar\delta(q+q')}{-i \omega + M(\bm{q})}\equiv G_0(q) \delta_{ab}
\bar\delta(q+q'),
\label{Eq:propagator}
\end{equation}
where $\langle \ldots \rangle_0$ stands for the average over noninteracting
theory~\eqref{Eq:noninter}, and 
the delta function $\bar\delta(q+q') := (2\pi)^{d+1} \delta(\omega + \omega')\delta(\bm{q} + \bm{q}')$.
Graphical representation of the propagator and the interaction terms of 
the action~\eqref{msr3} is shown in Figure~\ref{Fig:bb}. 

The generating functional of the correlation functions is
\begin{equation}
Z[J,\widetilde J] = \int {\cal D}\widetilde \phi {\cal D}\phi
\exp\left (-S + \int_x \widetilde J \cdot \widetilde \phi + J \cdot \phi \right ),
\end{equation}
where $J \cdot \phi = \sum_a J_a \phi_a$ and $\widetilde J \cdot \widetilde\phi = \sum_a \widetilde{J}_a \widetilde{\phi}_a$.
The cumulants can be calculated by functional derivative of 
$F[\widetilde J,J] = \ln Z[\widetilde J,J]$ with respective to the
sources such that
\begin{eqnarray}
G_{\tilde n,n} (q_1,\ldots,q_{\tilde n};p_{1},\ldots,p_n)\nonumber \\
= \left \langle \prod_{i=1}^{\tilde n} \widetilde \varphi_{a_i} (q_i)
\prod_{k=1}^{n} \widetilde \varphi_{b_k} (p_k) \right \rangle_c
\nonumber\\
= \left . 
\prod_{i=1}^{\tilde n}  \frac{\delta}{\delta \tilde j_{a_i}(-q_i)}
\prod_{k=1}^{n}  \frac{\delta}{\delta j_{b_k}(-p_k)}
F[\tilde j,j] \right |_{\tilde j = j = 0},
\end{eqnarray}
where $\tilde j$ and $j$ are the Fourier transformation of $\tilde J$ and
$J$, respectively, and the multiplication factor $(2\pi)^{d+1}$ is assumed
in the functional derivative with respective to 
$j$ or $\tilde j$. This convention will also be used in Eq.~\eqref{Eq:psi}. 
For convenience, the field indices are not written explicitly 
in $G_{\tilde n,n}$.
The vertex functions $\Gamma_{\tilde m,m}$ can be obtained from $G_{\tilde n,n}$
by a Legendre transformation
\begin{equation}
\Gamma[\widetilde \psi,\psi] = - F + \int_q 
\left ( \tilde j(-q) \cdot \widetilde \psi(q)
+j(-q) \cdot\psi(q) \right ),
\end{equation}
where
\begin{equation}
\widetilde \psi_a(q) = \frac{\delta F}{\delta \tilde j_a(-q)}, \quad
\psi_a(q) = \frac{\delta F}{\delta j_a(-q)}.
\label{Eq:psi}
\end{equation}

The fields are written in terms of renormalized fields as
\begin{eqnarray}
\widetilde \phi_a =  Z_{\widetilde \phi}^{1/2} \widetilde \phi_{aR},\;\;
\phi_a = Z_\phi^{1/2}  \phi_{aR}, \;\;
Z \equiv \sqrt{Z_{\widetilde \phi}Z_\phi},
\end{eqnarray}
and the parameters in terms of renormalized parameters as
\begin{eqnarray}
 D &=&  \frac{Z_D}{Z} D_R,\;\; \rho = \frac{Z_{\rho}}{Z_D} \rho_R,\;\;
r = \frac{Z_{r}}{Z_D} r_R \mu^2, \\ 
u_0 &=& \frac{Z_0}{ZZ_\phi} u_{0R},\;\;
u_1 = \frac{Z_1}{ZZ_\phi} u_{1R},\;\;
u_2 = \frac{Z_2}{ZZ_\phi} u_{2R},\nonumber \\
T &=& \frac{Z_T}{Z_{\widetilde \phi}} T_R,\;\;
e_p =  \frac{Z_p}{ZZ_\phi^{1/2}} e_{pR},\;\;
e_m = \frac{Z_m}{ZZ_\phi^{1/2}} e_{mR}, \nonumber
\end{eqnarray}
where $R$ in the subscripts stands for the renormalized quantities and
$\mu$ is an arbitrary momentum scale. 
Substituting these parameters in $\Gamma[\widetilde \psi,\psi]$ gives the generator $\Gamma_R[\widetilde \psi,\psi]$
of the renormalized vertex functions:
\begin{eqnarray}
\Gamma_{\tilde m,m}^{a_1\ldots a_{m+\tilde m}}(\{q_i\})=\nonumber\\
 \left . \prod_{i=1}^{\tilde m}  \frac{\delta}{\delta \widetilde \psi_{a_i}(-q_i)}
\prod_{j=\tilde m+1}^{\tilde m +m}  \frac{\delta}{\delta \psi_{a_j}(-q_j)}
\Gamma_R[\widetilde \psi,\psi] \right |_{\tilde \psi = \psi = 0}.
\end{eqnarray}

The renormalization factors are determined by the following set of normalization conditions:
\begin{widetext}
\begin{eqnarray}
\Gamma^{11}_{1,1}(0;0) = r_R \mu^2, \;\; 
\Gamma^{11}_{2,0}(q=0) = - 2 T_R, \;\; 
\Gamma_{1,3}^{1111}(q_i=0) = u_{0R},\;\;
\Gamma_{1,3}^{1122}(q_i=0) = u_{1R}, \;\;
\Gamma_{1,3}^{1133}(q_i=0) = u_{2R},\\
\left . \frac{\partial}{\partial (i \omega)} \Gamma^{11}_{1,1}(-q;q) \right |_{q=0}=1
\quad 
\left  .\frac{\partial}{\partial (\bm{q_\perp}^2)} \Gamma^{11}_{1,1}(-q;q) \right |_{q=0} =D_R, \quad 
\left . \frac{\partial}{\partial (q_\|^2)} \Gamma^{11}_{1,1}(-q;q) \right |_{q=0} = \rho_R,\nonumber \\
\left . \frac{\partial}{\partial (ik_\|)}\Gamma_{1,2}^{123}\left (-k;\frac{k}{2},\frac{k}{2}\right ) \right |_{k=0} = e_{pR},\nonumber \quad 
\left . \frac{\partial}{\partial (ik_\|)}\Gamma_{1,2}^{123}\left (0;-\frac{k}{2},\frac{k}{2}\right ) \right |_{k=0} = e_{mR},
\end{eqnarray}
where the momentum conservation for each vertex functions has already been 
taken into account (so no delta functions are multiplied).
Employing the dimensional regularization with minimal subtraction 
scheme~\cite{amit}, along with the normalization conditions above, we obtain the renormalization factors to one-loop order  (see Appendix for the details) as follows.
\begin{eqnarray}
Z = 1 + \frac{v_m^2}{4 \epsilon},\quad
Z_D = 1 + \frac{v_m^2}{6 \epsilon},\quad
Z_T = 1 - \frac{v_m^2}{2 \epsilon},\quad
Z_\rho = 1 + \frac{3}{4\epsilon} \left (v_m^2 - v_p^2  \right ), \label{Eq:Zfactors}\\
Z_r = 1 + \frac{1}{2\epsilon} \left ( g_0 + g_1 + g_2 + 2 v_m^2 \right ),\quad
Z_p = 1 + \frac{1}{8 \epsilon} \left (  7 (g_1 + g_2) -3 v_m^2 - 2 \frac{v_m}{v_p} (g_1 - g_2) \right ),\nonumber \\
Z_e = 1 +\frac{1}{8\epsilon} \left ( 3 v_m^2 + 2 v_p^2 + 3 \frac{v_p}{v_m}(g_1 - g_2) \right ),
\quad
Z_0 = 1+
\frac{3 g_0}{2 \epsilon} + \frac{3 g_1 g_2}{\epsilon g_0} 
+ \frac{3 v_m^2}{4 \epsilon g_0}( g_1 + g_2 - 2 v_m^2 + 2 v_p^2 ), 
\nonumber
\\
Z_1 = 1 + \frac{g_2^2}{2 g_1 \epsilon} + \frac{v_m^2}{4 g_1 \epsilon}
\left(g_0+g_1-2 \left(g_2+v_m^2\right)\right)+
 \frac{1}{\epsilon} ( g_0 + g_1+g_2) + \frac{v_m v_p}{4 g_1 \epsilon}  \left(g_1+g_2+v_m \left(3 v_m-v_p\right)\right),\nonumber \\
Z_2 = 1 + \frac{g_1^2}{2 g_2 \epsilon} + \frac{v_m^2}{4 g_2 \epsilon}
\left(g_0+g_2-2 \left(g_1+v_m^2\right)\right)+
 \frac{1}{\epsilon} ( g_0 + g_1+g_2) - \frac{v_m v_p}{4 g_2 \epsilon}  \left(g_1+g_2+v_m \left(3 v_m+v_p\right)\right),\nonumber 
\end{eqnarray}
where $\epsilon=4-d$;  and the dimensionless expansion parameters
\begin{eqnarray}
g_i = \frac{A_d T}{D^2 \sqrt{\rho}} u_i \mu^{-\epsilon},\;
v_s = \left ( \frac{A_d T}{D^3 \rho^{3/2}}\right )^{1/2} e_s \mu^{-\epsilon/2},
\label{Eq:parameters}
\end{eqnarray}
where, $i=0$, 1, or 2, and $s$  is either $p$ or $m$; and
the convenient geometric factor $A_d = 2 \Gamma(3-d/2)/(4 \pi)^{d/2}$, where $\Gamma$ here is the Euler gamma 
function. Furthermore, we obtain the following RG flow equations to one-loop order,
\begin{eqnarray}
\mu \frac{d g_0}{d \mu} &=& - g_0 \epsilon + \frac{3}{8}g_0(4g_0 + v_p^2) 
+ 3 g_1 g_2
+ \frac{v_m^2}{8} ( 6 (g_1 + g_2) - 11 g_0 + 12 v_p^2 - 12 v_m^2),\\
\mu \frac{d g_1}{d\mu} &=& g_1 \left(-\epsilon \!+\! g_0\!+\!g_1\!+\!g_2 \right)+\frac{g_2^2}{2}+
\frac{v_p^2}{8}  \left(3 g_1-2 v_m^2\right)+\frac{v_m v_p}{4} 
   \left(g_1+g_2+3 v_m^2\right)+\frac{v_m^2}{8}  \left(2 g_0\!-\!9 g_1\!-\!4 g_2\!-\!4 v_m^2\right),\\
\mu \frac{d g_2}{d\mu} &=& g_2 \left(-\epsilon \!+ \!g_0\!+\!g_2\!+\!g_1 \right)+\frac{g_1^2}{2}+
\frac{v_p^2}{8}  \left(3 g_2-2 v_m^2\right)-\frac{v_m v_p}{4} 
   \left(g_2+g_1+3 v_m^2\right)+\frac{v_m^2}{8}  \left(2 g_0\!-\!9 g_2\!-\!4 g_1\!-\!4 v_m^2\right),\\
 \mu \frac{d v_p}{d \mu}
&=& \frac{v_p}{16} (-8 \epsilon + 14 (g_1 + g_2) - 22 v_m^2 + 9 v_p^2 )
-\frac{v_m}{4}(g_1-g_2),\\
\mu \frac{d v_m}{d \mu}
&=& \frac{v_m}{16} \left ( - 8 \epsilon + 13 v_p^2 -10 v_m^2 \right )
+\frac{3}{8} v_p (g_1 - g_2).
\end{eqnarray}
\end{widetext}
 We solve for the RG fixed points numerically and find 72 fixed-point solutions. Out of these, 56 solutions are complex valued and, hence, being unphysical, are discarded, while the rest of the fixed points are discussed below.  We denote a fixed point as $( g_0^*, g_1^* , g_2^*,v_p^*, v_m^* )$. After identifying the fixed points, we analyze the linearized flow equations to find their (local) stability. 

There are four equilibrium fixed points, for which the bias couplings vanish,  
$v_p^* = v_m^*=0$,  and the other couplings are as follows:
\begin{subequations}
\label{Eq:fix_eq}
\begin{eqnarray}
\text{Gaussian:}~~ g_0^* &=& g_1^* =g_2^*= 0,   \\
\text{Ising:}~~ g_0^* &=& \frac{2}{3}  \epsilon, ~~ g_1^*=g_2^*=0, \\
\text{Cubic:}~~g_0^* &=& \frac{4}{9} \epsilon, ~~ g_1^* =g_2^* =\frac{2}{9} \epsilon, \\
\text{Heisenberg:}~~ g_0^* &=& \frac{6}{11}\epsilon , ~~ g_1^* =g_2^* =\frac{2}{11}\epsilon .
\end{eqnarray}
\end{subequations}
In the absence of spatial-anisotropic perturbations,
the Heisenberg fixed point is stable for  $\epsilon>0$ (i.e., $d < 4$), while the Gaussian fixed point is stable for $d>4$ .
Note that, even if the system has only the cyclic-permutation 
symmetry, the full permutation symmetry is restored asymptotically.

In addition to those listed in~\eqref{Eq:fix_eq},  we find the following fixed points that also respect the full permutation-symmetric theory (namely, $v_m = 0$ and $g_1=g_2$):
\begin{subequations}
\label{Eq:FPbias}
\begin{eqnarray}
\text{P}_\text{G}:~~g_0^* &=& g_1^* = 0, ~~E_p^* = \frac{8}{9} \epsilon,
\label{Eq:bias_fixed}
\\
\text{P}_\text{I}:~~g_0^* &=& \frac{4}{9} \epsilon,~~ g_1^* = 0, ~~ E_p^* = \frac{8}{9} \epsilon,
\label{Eq:ersatz_ising}\\
\text{P}_\text{C}:~~g_0^* &=& \frac{2}{5} \epsilon, ~~g_1^* = \frac{1}{5} \epsilon,~~ E_p^* = \frac{4}{15} \epsilon,\\
\text{P}_\text{H}:~~g_0^* &=& \frac{6}{13} \epsilon, ~~g_1^* = \frac{2}{13} \epsilon,
~~E_p^* = \frac{16}{39}\epsilon,
\label{Eq:stable}
\end{eqnarray}
\end{subequations}
where $E_p \equiv v_p^2$.  The fixed points are so labeled because of the structural similarity with the corresponding points in Eq.~\eqref{Eq:fix_eq}.  Here, we do not distinguish between $v_p^*$ and $-v_p^*$, since choosing one of them amounts to choosing the bias direction. More precisely, changing sign of bias parameters $\{ v_p, v_m \} \to \{ -v_p, -v_m \} $ will not take us to a new fixed point with different critical exponents.

When all the couplings are tuned off except $e_p$, RG flows only along the line $g_0=g_1=g_2=v_m=0$.
In this case, there are two fixed points, Gaussian and P$_\text{G}$, and the fixed point~P$_\text{G}$ 
is stable (unstable) if $\epsilon > 0$ ($\epsilon <0$).
To our knowledge, this fixed-point was not known before in the literature. In the one-component case, $N=1$,  a similar stable fixed point is found by Hwa and Kardar~\cite{hwa}.

For the choice $u_1=u_2=e_m=0$ in Eq.~\eqref{msr3}, the one-loop calculations show that there are 
four fixed points (Gaussian, Ising, P$_\text{G}$, and P$_\text{I}$),  where P$_\text{I}$ is the only stable fixed-point
for $\epsilon > 0$. Though the linear stability analysis alludes to the existence 
of a universality class in $(g_0,  E_p)$ subspace, the higher-order loop corrections rules out this possibility. 
For instance, adding a $\widetilde \phi_1 \phi_1^3$
vertex to Fig.~\ref{Fig:phi4}(c) with $a=1$, $b=c=c_2=c_3=2$, and $c_1 = 3$ ,
generates $g_1$ at two-loop order, and, hence, the RG flow pushes the system out of $(g_0,  E_p)$ subspace.

Thus, in the space of all perturbations that preserve the full permutation symmetry ($u_1 = u_2$ and
$e_m = 0$), there are eight fixed points,  among which only one is stable;
P$_\text{H}$ is stable for
$\epsilon >0$ ($d<4$),  while Gaussian is stable for $\epsilon < 0$ ($d>4$). 
Hence, unlike $N\!=\!1$ case, the bias perturbations in $N=3$ case are highly relevant and can lead to a new universality class.

Let us now also include $v_m$ term, which breaks the permutation symmetry to cyclic-permutation symmetry.
Suppose we first restrict to a subspace with the choice $v_p = 0$ and $g_1 = g_2$, then the action \eqref{msr3} is invariant
under the transformation $\phi_1 \leftrightarrow \phi_2$ and $e_m \rightarrow-e_m$, and hence the symmetry 
constraints the RG flow to the $(g_0,g_1=g_2,v_m)$-subspace.  For $\epsilon > 0$, the fixed points are the equilibrium ones with $v_m^*=0$. 
This special case was studied earlier in Ref.~\cite{jayajit}, where,
in contrast to our result, a stable fixed point was found.
However, in Ref.~\cite{jayajit} it was numerically observed that the system 
exhibits chaotic behavior in the noiseless (zero-temperature) limit. 
We argue that this numerical observation is more consistent with our 
result than the existence of a stable fixed point. The absence of a stable fixed point signifies that 
there is no order-disorder phase transition. 
Since the system in the infinite-temperature limit should be fully disordered 
and in the zero-temperature limit behavior is also chaotic, there is no ordered phase in the system (assuming there is at most one transition).
We thus expect that there is no phase transition, or in other words, no stable fixed point, as confirmed by our analysis.

If we do not restrict to $(g_0,g_1=g_2,v_m)$ subspace and instead explore the space of all coupling constants,  we then obtain the following fixed point:
\begin{eqnarray}
g_0^* = 1.49763 \epsilon , g_1^* = -1.86313 \epsilon , g_2^* = 1.12359 \epsilon , \nonumber\\
v_p^* =2.02811 \sqrt{\epsilon},
v_m^* =  1.06466 \sqrt{\epsilon},
\end{eqnarray}
which is unstable, and also find three other unstable fixed points.
The other unstable fixed points can be obtained from the above by taking  $\{ v_p^*, v_m^* \} \to \{ -v_p^*, -v_m^* \} $, and $\{ g_1^*, g_2^*, v_p^* \} \to \{ g_2^*, g_1^*, -v_p^* \}$, or 
$\{ g_1^*, g_2^*, v_m^* \} \to \{ g_2^*, g_1^*, -v_m^* \}$. 
Note that the flow equations are invariant under these transformations.
 
The fixed-point analysis tells us that the presence of $e_m$ term will destabilize the $N=3$ field theories.

\section{\label{Sec:sum}Summary}

To sum up, we have studied the effect of spatially anisotropic perturbations on nonconserved $N$-vector models.

We first constructed spatially anisotropic $N$-vector models that obey Langevin dynamics, and contain the most general marginal interactions 
at $d=4$.  If the dynamics is invariant under all the permutations of the 
field components, then the number of coupling constants can be at most 
12 (7 $\phi^4$-type $G$-couplings and 5 bias  ${\cal E}$-couplings). We then 
argued that single-length-scale field theories with bias are possible only for $N=1$ or certain $N=3$ models. 
The $N =1$ (BS) theory has been studied earlier~\cite{bassler}, where the bias was found to be marginally irrelevant.
For $N=2$ or $N > 3$, we see that the bias generates off-diagonal mass terms and rules out the possibility of Langevin field theories with a single length scale. 
Hence, the $N=2$ models and the generic $N>3$ models, when subjected to bias, should behave like the BS model~\cite{bassler} in the large-distance limit.  We also confirmed this by analyzing numerically an $N=2$ model with bias. 

For $N=3$ field theories with a single length scale, the full permutation symmetry allows only one kind of bias coupling (labeled $e_p$), while    
the cyclic-permutation symmetry allows another additional coupling (labeled $e_m$).We followed the renormalization-group (RG) flows, up to 
one-loop order, for the $N=3$ systems that are invariant under cyclic-permutations of the field components.  
In this case, the coupling-constant space is five dimensional, with three $\phi^4$-type couplings ($u_0, u_1, u_2$) and two bias 
couplings ($e_p$ and $e_m$); see Eq.~\eqref{msr3}. We find that, in the 
presence of $e_m$-perturbations, no stable fixed point exists.

Once the $e_m$ term is thrown away (and $u_1 = u_2$ is set), 
the system becomes permutation symmetric and has
eight fixed points as given in Eqs~\eqref{Eq:fix_eq} and  \eqref{Eq:FPbias}.
Only two of the fixed points are stable: the fixed-point P$_\text{H}$ [see Eq.~\eqref{Eq:stable}] is stable
for $\epsilon >0$ ($d<4$), while the Gaussian fixed point  is stable for
$\epsilon < 0$ ($d>4$).  Hence, we find a new universality class governed by the fixed point P$_\text{H}$ for $N=3$ systems with a 
spacial-anisotropic bias. 

We also find another universality class when all the coupling constants except $e_p$ are tuned off. In this case, the RG flow does not generate other couplings and leads to a nontrivial stable fixed point, denoted  P$_\text{G}$ [see Eq.~\eqref{Eq:bias_fixed}].

In general, nonconserved $N$-vector models are sensitive to 
spatial-anisotropic perturbations, and the large-distance properties are 
governed by the kinetic Ising class, except for $N=3$. In the case of $N=3$, 
we found two universality classes governed by P$_\text{G}$ and 
P$_\text{H}$.

\acknowledgments
SD acknowledges the generous support by the people of South Korea, and Korea Institute for Advanced Study, where the work was initiated.
S.-C.P. would like to acknowledge the support by DFG within SFB 680 \textit{Molecular Basis of Evolutionary Innovations};
the support by the Catholic University of Korea, Research Fund, 2010;
and the support by the Basic Science Research Program through the National Research Foundation of Korea (NRF) funded
by the Ministry of Education, Science and Technology (Grant No. 2010-0006306)

\appendix*

\section{One loop calculations}
\subsection{Integrals in dimensional regularization}
The list of integrals required for
dimensional regularization~\cite{amit}:
\begin{eqnarray}
\int \frac{d^d\bm{p}}{(2\pi)^d} \left ( \bm{p}_\perp^2 + \rho p_\|^2 + r\right )^{-2}
= \frac{A_d r^{-\epsilon/2}}{\epsilon \rho^{1/2}},\\
\int \frac{d^d\bm{p}}{(2\pi)^d} p_\|^2\left ( \bm{p}_\perp^2 + \rho p_\|^2 + r \right )^{-3}
= \frac{A_d r^{-\epsilon/2}}{4 \epsilon \rho^{3/2}} ,\\
\int \frac{d^d\bm{p}}{(2\pi)^d} p_\|^4\left ( \bm{p}_\perp^2 + \rho p_\|^2 + r \right )^{-4}
= \frac{A_d r^{-\epsilon/2}}{8 \epsilon \rho^{5/2}} ,\\
\int \frac{d^d\bm{p}}{(2\pi)^d} p_2^2 p_\|^2\left ( \bm{p}_\perp^2 + \rho p_\|^2 + r \right )^{-4}
= \frac{A_d r^{-\epsilon/2}}{24 \epsilon \rho^{3/2}} ,
\end{eqnarray}
where $p_2$ in the last equation stands for one of the perpendicular component
of $\bm{p}$ and $A_d \equiv 2 \Gamma(3 - d/2)/ (4 \pi)^{d/2}$ is a geometric factor.
\subsection{Diagrammatics}
Let $V_i$ denote the number of $i$-legged vertices  
(see. Fig.~\ref{Fig:bb})  in the loop expansion for
$\Gamma_{\tilde m,m}$, and $V = V_2+V_3+V_4$ denote the total number of vertices in a diagram.  Since the number of $\varphi$
fields in the internal integration should be equal to that of $\tilde \varphi$, the number of internal lines $I$ is given by
\begin{equation}
I \equiv 2 V_2 + V_3 + V_4 - \tilde m = 2 V_3 + 3 V_4 - m .
\end{equation}
If there are $L$ number of loops, then the relation $I-V = L-1$ should hold, and it therefore follows that,
\begin{equation}
V_2 = \tilde m + L-1,\quad
V_3 + 2 V_4 = m + \tilde m + 2(L-1),
\end{equation}
which, in the case of one-loop calculations, reduce to
\begin{equation}
V_2 = \tilde m,\quad V_3 + 2 V_4 = m + \tilde m.
\label{Eq:loop-vertex}
\end{equation}
\begin{figure}[t]
\includegraphics[width=0.45\textwidth]{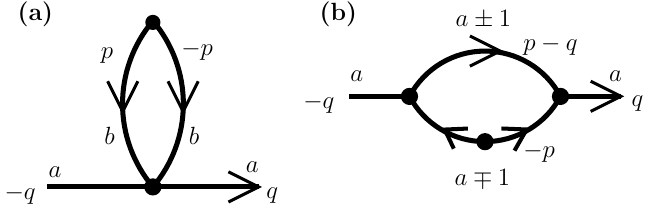}
\caption{\label{Fig:mass} One-loop diagrams for $\Gamma_{1,1}^{aa}(-q;q)$.
The internal momentum which should be integrated out is denoted by $p$. The value of $b$ can be 1, 2 or 3.}
\end{figure}
 
For notational convenience, we define the following function that is associated to momentum-dependent 3-legged vertices, 
\begin{eqnarray}
\lambda(q_1,q_2,q_3) = i e_p q_{1\|} + i e_m (q_{2\|}-q_{3\|}),
\end{eqnarray}
and the correlation function, 
\begin{equation}
C_0(p) = G_0(p) G_0(-p).
\end{equation}
\subsection{$\Gamma_{1,1}^{aa}(-q;q)$}
The solutions of Eq.~\eqref{Eq:loop-vertex} for $\tilde m = m = 1$ are
(a) $V_3 = 0$ and $V_4 = 1$ and (b) $V_3 = 2$ and $V_4=0$.
The corresponding one-particle irreducible (1PI) diagrams 
are illustrated in Fig.~\ref{Fig:mass}.
Note that none of the  diagrams in Fig.~\ref{Fig:mass}
can generate off-diagonal mass terms, as expected from our general arguments.
The loop integrals for Fig.~\ref{Fig:mass} are
\begin{figure}[t]
\includegraphics[width=0.22\textwidth]{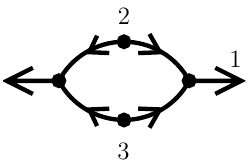}
\caption{\label{Fig:noise} One-loop diagram for 
$\Gamma_{2,0}^{11}(-q,q)$.}
\end{figure}
\begin{widetext}
\begin{eqnarray}
\textrm{(a)} &=& T (u + u_1 + u_2) \int_p C_0(p)
= - D r \frac{u+u_1+u_2}{2} B_\epsilon,\\
\textrm{(b)}&=& -2 T \int_p G_0(p-q)C_0(p)
\left [ \lambda(q,p-q,-p)\lambda(-p+q,p,-q)
+ \lambda(q,-p,p-q) \lambda(-p+q,-q,p) \right ]\nonumber\\
&=&- \left ( i \omega e_m^2 + D \bm{q}_\perp^2 \frac{e_m^2}{6} +
D \rho q_\|^2 \frac{3}{4} \left(e_m^2-e_p^2\right) + D r
e_m^2 \right ) C_\epsilon
+ \ldots,
\end{eqnarray}
\end{widetext}
where 
\begin{eqnarray}
B_\epsilon = \frac{A_d T}{D^2 \sqrt{\rho} \epsilon}
r^{-\epsilon/2},\quad
C_\epsilon =  \frac{A_d T}{D^3 \rho^{3/2} \epsilon}
r^{-\epsilon/2},
\end{eqnarray}
and the ellipsis contains the finite parts that shall be dropped in the minimal-subtraction (MS) scheme that we adopt.
In the remainder, the integrals are evaluated in the MS scheme.
\subsection{$\Gamma_{2,0}^{11}(-q,q)$}
There are two solutions for Eq.~\eqref{Eq:loop-vertex} 
with $m=0$ and $\tilde m=2$:  \{$V_3=2$, $V_4=0$\} and \{$V_3 = 0$, $V_4=1$\}.
However, the later solution does not yield any 1PI
diagram.
Hence, there is only one diagram which is depicted in Fig.~\ref{Fig:noise}.
 The one-loop correction to $\Gamma_{2,0}^{11}(-q,q)$ is
\begin{eqnarray}
-4 T^2 \int_p C_0(p)^2 | \lambda(0,p,-p)|^2
= -  T e_m^2 C_\epsilon.
\end{eqnarray} 
\subsection{$\Gamma_{1,2}^{123}(q_1;q_2,q_3)$}
\begin{figure}[b]
\includegraphics[width=0.45\textwidth]{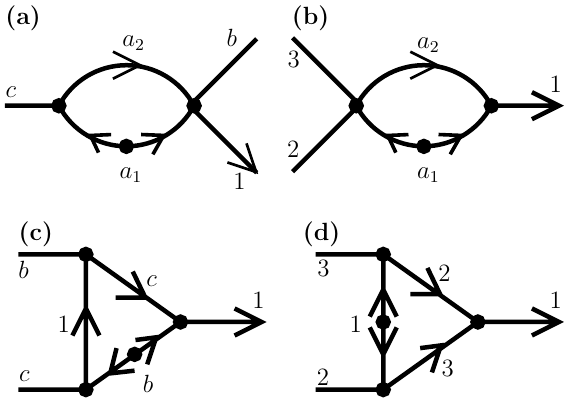}
\caption{\label{Fig:bias} One-loop diagrams for 
$\Gamma_{1,2}^{123}(q_1;q_2,q_3)$. For the external line with index $i$, 
put the momentum $-q_i$ and use the momentum conservation at each vertex point.
(a) If $a_1=1$ ($a_2=1$), then $a_2=b$
($a_1=b$). Since $b$ can be either 2 or 3, there are four different
diagrams with this form. (b) Either $\{a_1=2,a_2=3\}$ or $\{a_1=3,a_2=2\}$.
(c) Either $\{b=2,c=3\}$ or $\{b=3,c=2\}$. (d) There is a unique
diagram.}
\end{figure}

When $\tilde m =1$ and $m=2$, there are two solutions for
Eq.~\eqref{Eq:loop-vertex}: \{$V_3=1$, $V_4=1$\} or \{$V_3=3$, $V_4=0$\}.
For each set of solutions, two different 1PI diagrams can be drawn. 
For $V_2=V_3=V_4=1$, the diagrams are given in Fig.~\ref{Fig:bias}~(a) and (b),
while the diagrams for the other solution are given in Fig.~\ref{Fig:bias}~(c) and (d).
We now calculate these diagrams one by one.

For Fig.~\ref{Fig:bias} (a), depending on the values of
$\{b,c,a_1,a_2\}$, four different combinations are possible:
(a1) $b=2$ and $c=3$, and either \{$a_1 = 1$, $a_2=2$\} or \{$a_1=2$, $a_2=1$\};
and (a2) $b=3$ and $c=2$, and
either \{$a_1 = 1$, $a_2=3$\} or \{$a_1=3$, $a_2=1$\}.
Let us define the following operation ${\cal O}$, which transforms the momenta and coupling constants of the diagram: 
\begin{equation}
{\cal O} = \{ q_2\leftrightarrow q_3, \quad u_1\leftrightarrow u_2, \quad
e_m \rightarrow -e_m\}.
\end{equation}
Due to the cyclic symmetry, the result for (a2) can be readily acheived by operating
${\cal O}$ to (a1).
Since
\begin{eqnarray}
\textrm{(a1)} = - 2 T u_1 \int_p G_0(p-q_3) C_0(p) \times\nonumber \\
\times \left ( \lambda(-p+q_3,-q_3,p) + \lambda(-p+q_3,p,-q_3)
\right )\nonumber \\
= -\frac{3}{4} i e_p q_{3\|} u_1 B_\epsilon ,
\end{eqnarray}
we get
\begin{equation}
\textrm{(a)} = -\frac{3}{4} i e_p B_\epsilon \left ( q_{3\|} u_1 
+ q_{2\|} u_2 \right ) .
\end{equation}

For Fig.~\ref{Fig:bias} (b), either \{$a_1=2$, $a_2=3$\} or 
\{$a_1=3$, $a_2=2$\} should be satisfied. Note that the
interaction strength for $\widetilde \varphi_2 \varphi_2 \varphi_3^2$
is $u_1$ and that for $\widetilde \varphi_3 \varphi_3 \varphi_2^2$
is $u_2$. Hence, we get
\begin{eqnarray}
&&(b) = - 2 T\int_p G_0(p+q) C_0(p) \times\nonumber\\
&&\times 
\left ( u_1 \lambda(-q_1,p+q_1,-p) + u_2 \lambda(-q_1,-p,p+q_1)
\right )
\nonumber\\
&&=-i q_{1\|} \left(\left(u_1-u_2\right) e_m-2
   \left(u_1+u_2\right) e_p\right)\frac{ B_\epsilon}{4} .
\end{eqnarray}

For Fig.~\ref{Fig:bias} (c) and (d), it is convenient to introduce
\begin{eqnarray}
I_d(p;q_1,q_2) = C_0(p) G_0(p-q_2) G_0(p+q_1),\\
I_e(p;q_2,q_3) = C_0(p) G_0(p-q_2) G_0(-p-q_3).
\end{eqnarray}
It is easy to see that either \{$b=2$, $c=3$\} or 
\{$b=3$, $c=2$\} should be satisfied in Fig.~\ref{Fig:bias} (c). Thus, 
\begin{widetext}
\begin{eqnarray}
\textrm{(c)}=
2 T \int_p I_d(p;q_1,q_3) \lambda(-q_1,-p,p+q_1) \lambda(-p-q_1,p-q_3,-q_2)
\lambda(-p+q_3,p,-q_3)\nonumber\\
+
2 T \int_p I_d(p;q_1,q_2) \lambda(-q_1,p+q_1,-p) \lambda(-p-q_1,-q_3,p-q_2)
\lambda(-p+q_2,-q_2,p)\nonumber\\
=- i e_p q_{1\|} C_\epsilon \frac{e_m^2 - e_p^2}{8}
+ i e_m q_{2\|} C_\epsilon \frac{(e_m - e_p)^2}{16}
- i e_m q_{3\|} C_\epsilon \frac{(e_m + e_p)^2}{16}.
\end{eqnarray}
For Fig.~\ref{Fig:bias} (d), we get
\begin{eqnarray}
\textrm{(d)} = 2 T \int_p I_e(p;q_2,q_3) \lambda(-q_1,-p-q_3,p-q_2) 
\lambda(-p+q_2,p,-q_2) \lambda(p+q_3,-q_3,-p)\nonumber\\
=- i e_p q_{1\|} C_\epsilon \frac{e_p^2 - e_m^2}{8}
+ i e_m q_{2\|} \frac{1}{16} \left(e_m+e_p\right) \left(5 e_m+3 e_p\right) C_\epsilon
- i e_m q_{3\|} \frac{1}{16} \left(e_m-e_p\right) \left(5 e_m-3 e_p\right) C_\epsilon .
\end{eqnarray}
\begin{figure*}[t]
\includegraphics[width=0.75\textwidth]{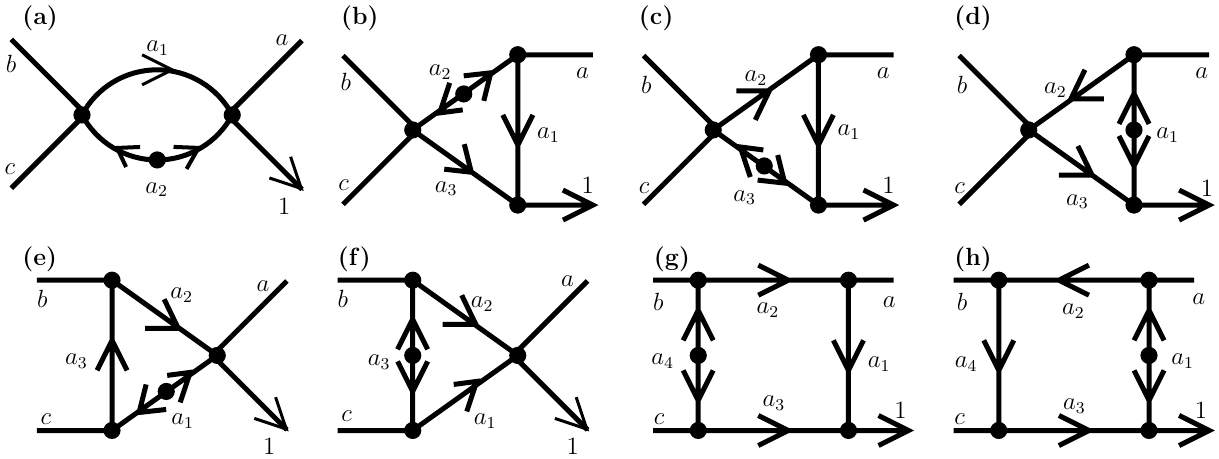}
\caption{\label{Fig:phi4} One-loop diagrams for 
$\Gamma_{1,3}^{1abc}(q_1;q_2,q_3,q_4)$.}
\end{figure*}
\subsection{$\Gamma_{1,3}^{1111}(q_1;q_2,q_3,q_4)$}
For $\tilde m = 1$, $m=3$, there are three solutions for Eq.~\eqref{Eq:loop-vertex}: $\{V_3 = 0, V_4=2\}$, $\{V_3=2, V_4=1\}$, or $\{V_3=4, V_4=0\}$.
There are eight different types of 1PI diagrams as shown in
Fig.~\ref{Fig:phi4}. Although the basic structure of diagrams is same for
any $\Gamma_{1,3}$'s, the mathematical expression for $\Gamma_{1,3}^{1111}$
is different from $\Gamma_{1,3}^{1122}(q_1;q_2,q_3,q_4)$.  Hence, we evaluate them separately; 
$\Gamma_{1,3}^{1111}$ is done in this section,  while 
$\Gamma_{1,3}^{1122}$ is done in the next. Note that $\Gamma_{1,3}^{1133}$ can be
 found easily by applying ${\cal O}$ to $\Gamma_{1,3}^{1122}$.
As a result of the normalization conditions, we will set $q_i=0$ from now on.

The diagrams for $\Gamma_{1,3}^{1111}$, in Fig.~\ref{Fig:phi4}  satisfy the condition $b=c=d=1$.  For the diagram in Fig.~\ref{Fig:phi4} (a), $a_1$ should be equal to $a_2$, and can take the values 1, 2 or 3. Hence the contribution from this diagram 
\begin{eqnarray}
\textrm{(a)} = - 6 T (u_0^2 + 2 u_1 u_2) \int_p G_0(p) C_0(p)
= -\frac{3}{2} \left ( u_0^2 + 2 u_1 u_2 \right ) B_\epsilon.
\end{eqnarray}

For Fig.~\ref{Fig:phi4} (b), (c) and (d), either \{$a_1 = 2$, $a_2=a_3=3$\} or 
\{$a_1=3$, $a_2=a_3=2$\} should be satisfied. Hence,
\begin{eqnarray}
\textrm{(b)} &=& 6 T \int_p \left \{ u_2 \lambda(0,p,-p) \lambda(p,0,-p)
+ u_1 \lambda(0,-p,p) \lambda(p,-p,0) \right\} I_{e0}(p)\nonumber\\
&=& -\frac{3}{4} e_m ((u_1+u_2)
   e_m+(u_2-u_1) e_p) C_\epsilon,\\
\textrm{(c)} &=& 6 T \int_p \left \{ u_2 \lambda(0,p,-p) \lambda(p,0,-p)
+ u_1 \lambda(0,-p,p) \lambda(p,-p,0) \right\} I_{d0}(p)
= \frac{1}{2} \textrm{(b)},\\
\textrm{(d)} &=& 6 T \int_p \left \{ u_2 \lambda(0,p,-p) \lambda(-p,p,0)
+ u_1 \lambda(0,-p,p) \lambda(-p,0,p) \right\} I_{d0}(p)
= \frac{1}{2} \textrm{(b)} -\frac{3}{4} \left(u_1-u_2\right) e_m e_p C_\epsilon,
\end{eqnarray}
where $I_{e0}(p) = I_{e}(p;0,0)$ and $I_{d0}(p) = I_{d}(p;0,0)$.

For Fig.~\ref{Fig:phi4} (e) and (f), either \{$a_3 = 2$, $a_2=a_1=3$\} or 
\{$a_3=3$, $a_2=a_1=2$\} should be satisfied. Hence,
\begin{eqnarray}
\textrm{(e)} &=& 12 T \int_p \left \{ u_2 \lambda(-p,p,0) \lambda(-p,0,p)
+ u_1 \lambda(-p,0,p) \lambda(-p,p,0) \right\} I_{d0}(p)
=\frac{3}{8} \left(u_1+u_2\right) \left(e_m^2-e_p^2\right)C_\epsilon,\\
\textrm{(f)} &=& 6 T \int_p \left \{ u_2 \lambda(p,0,-p) \lambda(-p,0,p)
+ u_1 \lambda(-p,p,0) \lambda(p,-p,0) \right\} I_{e0}(p)\nonumber\\
&=&  \frac{3}{8} \left(2 \left(u_2-u_1\right) e_m
   e_p+\left(u_1+u_2\right) e_m^2+\left(u_1+u_2\right)
   e_p^2\right) C_\epsilon
\end{eqnarray}

For Fig.~\ref{Fig:phi4} (g) and (h), we introduce, for notational convenience,
\begin{eqnarray}
I_f(p) \equiv G_0(p)^3 G_0(-p)^2,\quad
I_g(p) \equiv G_0(p)^4 G_0(-p).
\end{eqnarray}
The nonvanishing contribution occurs when either
\{$a_1=a_4 = 2$, $a_2=a_3=3$\} or
\{$a_1=a_4 = 3$, $a_2=a_3=2$\}.  Hence we obtain,
\begin{eqnarray}
\textrm{(g)} &=& -12 T \int_p I_f(p) \lambda(-p,p,0) \lambda(-p,0,p) 
\left \{ \lambda(p,0,-p)\lambda(0,p,-p) + \lambda(p,-p,0) \lambda(0,-p,p)
\right \} \nonumber \\ &=& 9 e_m^2 (e_m^2 - e_p^2) D_\epsilon,\\
\textrm{(h)} &=& -12 T \int_p I_g(p) \lambda(-p,p,0) \lambda(-p,0,p) 
\left \{ \lambda(-p,0,p)\lambda(0,-p,p) + \lambda(-p,p,0) \lambda(0,p,-p)
\right \} = \frac{1}{3}\textrm{(g)},
\end{eqnarray}
where $D_\epsilon = C_\epsilon/(8 D \rho)$.

\subsection{$\Gamma_{1,3}^{1122}(q_1;q_2,q_3,q_4)$}
For Fig.~\ref{Fig:phi4} (a), the loop integral is always $\int_p G_0(p) C_0(p)$.
We therefore have to decide which interaction terms are involved in the diagrams.
If $a=1$ and $b=c=2$, then $a_1$ equals to $a_2$ and can take any
index. If $a=b=2$ and $c=1$, then either \{$a_1=1$, $a_2=2$\} or
\{$a_1=2$, $a_2=1$\} should be satisfied.
Hence,
\begin{equation}
(a) = - \left ( u_1^2+u_0 u_1+u_2 u_1+\frac{u_2^2}{2}\right )
B_\epsilon.
\end{equation}

For Figs.~\ref{Fig:phi4} (b), (c) and  (d), either \{ $a=1$ and $b=c=2$ \} or
\{ $c=1$ and $a=b=2$\} is required. If $a=1$ and $b=c=2$, then either
\{ $a_1=2$, $a_2=a_3=3$ \} or \{ $a_1=3$, $a_2=a_3=2$\} should be satisfied. 
If $c=1$ and $b=a=2$, then it follows that $a_1=3$, $a_2=1$, $a_3 = 2$.
Hence,
\begin{eqnarray}
\textrm{(b)} &=& 2 T \int_p I_{e0}(p) \left \{
u_2 \lambda(p,-p,0) \left ( \lambda(0,-p,p) + 2 \lambda(0,p,-p)\right )
+ u_0 \lambda(0,p,-p) \lambda(p,0,-p)\right \}\nonumber\\
&=&-\frac{1}{4} e_m \left(\left(u_0-u_2\right)
   e_m+\left(u_0+u_2\right) e_p\right) C_\epsilon,\\
\textrm{(c)} &=& 2 T \int_p I_{d0}(p) \left \{
u_2 \lambda(p,-p,0)  \lambda(0,-p,p) + 2 u_1 \lambda(p,-p,0) \lambda(0,p,-p)
+ u_0 \lambda(0,p,-p) \lambda(p,0,-p)\right \}\nonumber \\
&=& -\frac{3}{8} e_m \left(\left(u_1+u_2\right)
   e_m+\left(u_2-u_1\right) e_p\right) C_\epsilon,\\
\textrm{(d)} &=& 2 T \int_p I_{d0}(p) \left \{
u_2 \lambda(-p,0,p) \left ( \lambda(0,-p,p) + 2 \lambda(0,p,-p)\right )
+ u_0 \lambda(0,p,-p) \lambda(-p,p,0)\right \}\nonumber\\
&=& -\frac{1}{8} e_m \left(\left(u_0-u_2\right)
   e_m-\left(u_0+u_2\right) e_p\right) C_\epsilon.
\end{eqnarray}

For Fig.~\ref{Fig:phi4} (e), there are three possible cases:
$\{a=1 ,b=c=2\}$, $\{b=1, a=c=2\}$, $\{c=1, a=b=2\}$. 
If $a=1$ and $b=c=2$, then either \{$a_1=a_2=1$, $a_3=3$\} 
or \{$a_1=a_2=3$, $a_3=1$\} should be satisfied.
If $c=1$ and $b=a=2$, then it follows that $a_1=2$, $a_2=1$, $a_3 = 3$. 
If $b=1$ and $c=a=2$, then it follows that $a_1=1$, $a_2=2$, $a_3 = 3$. 
 Hence, we obtain
\begin{eqnarray}
\textrm{(e)} &=& 4 T \int_p I_{d0}(p) 
\left \{ \lambda(-p,0,p) \left [ u_1 \lambda(-p,0,p) + u_2 
\lambda(-p,p,0) + u_0 \lambda(-p,p,0) \right ] + u_1
\lambda(-p,p,0)^2 \right \}\nonumber \\
&=& \frac{3}{8} \left(u_1+u_2\right) \left(e_m^2-e_p^2\right) C_\epsilon.
\end{eqnarray}

For Fig.~\ref{Fig:phi4} (f),  out of the three possible cases that appear in the case (e), 
$\{a=1 ,b=c=2\}$, $\{b=1, a=c=2\}$, $\{c=1, a=b=2\}$,  the last two cases are identical, and
it is therefore sufficient to  consider only two possibilities.
In the case \{$a=1$ and $b=c=2$\}, either \{$a_1=a_2=1$, $a_3=3$\} 
or \{$a_1=a_2=3$, $a_3=1$\} should be satisfied.
In the case \{$c=1$ and $b=a=2$\}, it follows that $a_1=2$, $a_2=1$, $a_3 = 3$.  We thus obtain
\begin{eqnarray}
\textrm{(f)} = 2 T \int_p I_{e0}(p) \left \{
\lambda(-p,0,p) \left [ 2 u_1 \lambda(p,-p,0) + u_0 \lambda(p,0,-p) 
\right ] + \lambda(p,-p,0) \lambda(-p,p,0) \right \}\nonumber \\
= \frac{1}{8} \left(2 \left(u_0-u_2\right) e_m e_p+\left(u_0-2
   u_1+u_2\right) e_m^2+\left(u_0+2 u_1+u_2\right) e_p^2\right) C_\epsilon.
\end{eqnarray}

Both Fig.~\ref{Fig:phi4} (g) and (h),  have two possibilities:
either  \{$a=a_4 = 1$, $b=c=a_1=2$, $a_2=a_3=3$\} or
\{$a=b=a_3=2$, $c=a_2=1$, $a_1=a_4=3$\}. Hence we obtain,
\begin{eqnarray}
\textrm{(g)} &=& - 4 T \int_p I_f(p) \lambda(-p,p,0) \lambda(p,-p,0)
\left ( \lambda(0,p,-p) \lambda(-p,p,0) + \lambda(0,-p,p) \lambda(-p,0,p)
\right ) \nonumber \\ &=& 3 e_m^2 \left(e_m-e_p\right)^2 D_\epsilon,\\
\textrm{(h)} &=& - 4 T \int_p I_g(p) \lambda(-p,p,0) \lambda(-p,0,p)
\left ( \lambda(0,p,-p) \lambda(-p,p,0) + \lambda(0,-p,p) \lambda(-p,0,p)
\right ) \nonumber \\ &=&e_m^2 \left(e_m^2-e_p^2\right) D_\epsilon.
\end{eqnarray}
\end{widetext}

\end{document}